\documentclass[12pt,preprint]{aastex}  
\def\alwaysmath#1{\ifmmode{#1}\else{$#1$}\fi}  

\slugcomment{Submitted to ApJ}  
  
\shorttitle{An empirical calibration of ${\alpha}$ }   
\shortauthors{Ferraro et al.}  
  
\begin{document}

\title{An empirical calibration of the mixing\--length parameter ${\alpha}$
\footnote{Based on  observations
collected at the European Southern Observatory (ESO), La Silla, Chile.
Also based on observations made with the Italian 
Telescopio Nazionale Galileo (TNG) operated on 
the island of La Palma by the Fundacion Galileo Galilei of 
the INAF (Istituto Nazionale di Astrofisica) at the Spanish Observatorio 
del Roque de los Muchachos of the Instituto de Astrofisica de Canarias.
}}  
  
\author{Francesco R. Ferraro\altaffilmark{}  
\footnote{Dipartimento di Astronomia Universit\`a   
di Bologna, via Ranzani 1, I--40127 Bologna, Italy,  
francesco.ferraro3@unibo.it},   
Elena Valenti\altaffilmark{2,}\footnote{INAF-Osservatorio   
Astronomico di  Bologna,  
 via Ranzani 1, I--40127 Bologna,  Italy},  
 Oscar Straniero\altaffilmark{}\footnote{INAF-Osservatorio 
 Astronomico di Collurania, 64100, Teramo, Italy}, 
  Livia Origlia\altaffilmark{3}  
   }
   
\begin{abstract}  
We present an empirical calibration of the Mixing\--Length
free parameter ${\alpha}$ based on a homogeneous Infrared
database of 28 Galactic globular clusters spanning a wide
metallicity range (-2.15$<$[Fe/H]$<$-0.2).  Empirical
estimates of the red giant effective temperatures 
have been obtained from infrared colors.
Suitable relations linking these temperatures to the cluster 
metallicity have been obtained and 
compared to theoretical predictions.
An appropriate set of models
for the Sun and  Population II giants have been
computed by using both the standard
solar metallicity $[Z/X]_{\odot}=0.0275$ and the most recently proposed
value $[Z/X]_{\odot}=$0.0177.
We find that when the
standard  solar metallicity   is adopted,  a unique value
of $\alpha$=2.17 can be used to reproduce both the solar
radius and the population II  red giant temperature.
Conversely, when the new
solar metallicity is adopted,  two  different values of $\alpha$ 
are required:  $\alpha=1.86$ to
fit the solar radius and  $\alpha \approx2.0$ to fit 
the red giant  temperatures.  
However, it
must be noted that, regardless the adopted solar reference, 
the $\alpha$ parameter does not show any significant dependence on metallicity.
\end{abstract}  
   
   
\section{Introduction}
\label{intro}
Stellar evolutionary models are common ingredients in a
variety of studies addressing fundamental cosmological and
astrophysical problems, such as  ages, formation processes,
and evolution of galaxies. However,  stellar sequences need
to be properly checked and calibrated before using them  to
derive properties of complex stellar systems. As
extensively reviewed by \cite{rf88}, beside model testing,
one should also take into account as a separated issue the
calibration of those quantities that theoretical models are
forced to parameterize because of the insufficient
understanding of the physical process. In this contest, the
lack of a rigorous theory of convection remains one of the
major deficiencies in the calculations of stellar
evolutionary sequences. Despite many efforts to replace the
mixing--length (ML) theory of convection by less crude 
approximations \citep[see][]{cm91,sp97,lud99},  the
classical formulation of \cite{bv58} continues to be 
universally used. The ML algorithm requires the presence of
the free parameter ${\alpha}=l/H_p$, which represents the
ratio between the  mean free path of a convective element
($l$) and the pressure scale height ($H_p$): the 
variations of this parameter strongly affects the structure
of the outer envelope (i.e. radius and temperature).  In
fact, this parameter determines the efficiency of energy
transport by convection in the outermost layer of a star:
for a  given stellar luminosity, it fixes the radius of the
star, hence  its temperature and colors. Evolutionary
tracks must then be calibrated by comparison with   stellar
radii and/or temperatures derived from observations. The
most obvious ML calibrator is the Sun.  The ML parameter
can be fixed by constraining theoretical solar models (i.e.
with solar mass, age and chemical composition) to reproduce
the solar radius \citep[see e.g.][]{m79,sw83,vdb83}. The
solar calibrations suggest that the parameter ${\alpha}$
should range from 1.5 to 2, when the classical ML algorithm
is used \citep{cg68}.  Observational evidences have shown
that similar values can account for  {\it (i)} the lower
Main--Sequence (MS) slope of young open clusters 
\citep{vdbb84}, {\it (ii)}  the positions in the
Color-Magnitude Diagram (CMD) of the local  Population II
sub-dwarfs \citep{buon88}, {\it (iii)} the properties of 
well--observed binaries whose components are in widely
separated evolutionary  phases \citep{vdbh85}, and {\it
(iv)} the effective temperature of  Red Giant Branch (RGB)
stars \citep{sc91,css95,vdb00,al99,al00}.  However,
although similar ${\alpha}$ values have been derived by
using different ML--calibrators, there is no theoretical 
justification that the same value of ${\alpha}$ should
apply to any star.

In addition, \cite{css95} emphasized that in the case of low\--mass stellar
models the derived stellar radii depend on the opacity which significantly 
contributes in determining the temperature gradient in their turbulent external 
layers.  Hence, stellar models, based on different opacity tables, could
require different values of ${\alpha}$. 

An homogeneous dataset of stellar temperatures at different metallicities 
to properly calibrate the ML parameter is urgently needed
before any further attempt to use evolutionary models
to derive relevant properties of stellar populations.

In this paper we present an empirical calibration of the
$\alpha$ parameter, based on accurate estimates of the RGB
effective temperature   at fixed luminosity levels for a
homogeneous sample of Galactic  Globular Clusters (GCs)
with  different metallicity,  observed by our group  in the
last 10 years 
\citep[see i.e.][]{f00,v04a,v04,bb04,v05}.
This infrared data\--set  
is the ideal tool to properly calibrate the ${\alpha}$ parameter
for low\--mass stellar models and to study its possible dependence on 
metallicity.

\section{The mixing length}

One of the main source of uncertainty in stellar models 
concerns the efficiency of convection. In the
external layer of a red giant star, the convective heat
transfer may substantially departs from the adiabatic
regime, in which any internal excess of heat is efficiently
transported outward by the ascending convective bubbles. In
the framework of the standard ML theory, the
convective efficiency is tuned by changing the $\alpha$
parameter. 
Large values of $\alpha$ (i.e. larger than 2.5) corresponds
to a very efficient convection. In this case, the
temperature gradient is close to the minimum allowed value
(the adiabatic one) and the predicted red giant temperature
are larger. On the contrary, for lower values of the
$\alpha$ parameter, the internal heat excess is only
partially removed by convection, and the radiative energy
transport plays an important role. This is the case of GC
RGB stars, where the stellar effective temperature is
sensitive to the radiative opacity. For this reason, models
computed with the same $\alpha$ value, but larger opacity,
produce lower effective temperatures. Hence a proper variation of 
$\alpha$ may counterbalance a variation of low temperature opacities
\citep{css95}. 

The $\alpha$ parameter is usually calibrated by forcing 
stellar models of the Sun to fit the
solar radius, which is known with high precision. 
However, 
the envelope of a red giant has rather different physical
characteristics with respect to
the solar envelope. A direct comparison  of
empirical temperatures of red giant stars
with the corresponding model predictions, 
provides an independent calibration of $\alpha$ and
allows us to measure the efficiency of the
convective energy transport in very different physical
conditions. 

\section{The empirical database}

The homogeneous data\--set of fiducial RGB ridge lines
presented here  is based on high quality  J, H and K
photometry of 28 Galactic GCs  published by our group in
the last few years \citep{f00,v04,v04a,bb04,v05}.  The data
were obtained at ESO La Silla Observatory (Chile) using 
IRAC2@ESO/MPI2.2m and SofI@NTT/ESO IR cameras  and at the
Telescopio Nazionale Galileo with ARNICA in several
observing runs.  On average, the central ${\sim}20$
arcmin$^2$ of each cluster has been mapped, allowing us to
sample a significant fraction of the total cluster light
(typically ${\approx}$70--90\%). For all the programme
clusters the same data reduction procedure has been applied
\citep[see][ for more details]{v04a}; the instrumental
magnitudes have been calibrated into the 2MASS photometric 
and astrometric system, allowing us to build the largest
homogeneous IR database of GCs ever obtained\footnote{The
photometric catalogs are  available in electronic form at
the CDS web site.}.

For all the clusters listed in Table~\ref{tab}, the observations
were deep enough to properly sample the entire RGB extension, from the base 
(typically 2--3 magnitudes below the Horizontal Branch (HB)) up
to the RGB\--Tip, thus allowing us a complete study of the RGB morphological
features and a clear definition of the mean ridge line
\citep[see e.g. Fig. 1 and 2 of][]{v04a,v05}.

The detailed procedure followed to obtain the RGB fiducial
ridge lines of the  clusters and to transform them into the
absolute plane  can be found in \citet{f00}.  Since for the
aim of this study the homogeneity of the data\--set is a
crucial issue,  we adopt the distance scale established by
\cite{f99} based on an empirical  measurement of the Zero
Age HB level in a sample of 61 Galactic GCs.
Note that the \cite{f99}   distance scale has been adopted by Ferraro et
al (2000), Valenti, Ferraro \& Origlia (2004b) in order to
perform a detailed comparison of the observed RGB Bump and Tip
luminosity   with theoretical expectations,
 finding an excellent agreement. However it is worth of
 noticing that the assumption of a different distance scale has
 a little effect on the derived  temperatures. In
 fact, even a significant difference in
 the adopted distance moduli of $\pm 0.1 mag$ would produce
 only  a
 difference of $\pm 30 K$ in the derived temperatures.
 
 Reddening
values have been taken from \cite{harris} for all the
clusters but the most extincted ones towards the Bulge
direction, for which a differential method  based on the
comparison of CMDs and Luminosity Functions (LFs) with
those of a reference  cluster of similar metallicity, has
been applied   \citep[see][ for more details]{v04a}. In
order to use these empirical data\--set for the calibration
of  theoretical models, the mean ridge lines in the $\rm
[M_K,(J-K)_0]$ observational plane must be transformed into
the M$_{Bol}$, Log~T$_e$ theoretical one.  In doing this we
adopted the bolometric correction (BC$_K$) and the
temperature scale computed by \cite{pm98}. These empirical
relations have been specifically calibrated  on
Population~II giants in Galactic GCs \citep[see
also][]{al99}. Fig.~\ref{lines} shows the fiducial RGB
ridge lines for the 28 Galactic  GCs in our sample in the
M$_{Bol}$, log(T$_e$) theoretical plane.  

\subsection{The metallicity scale}

As widely discussed in \cite{f99,f00} the location of the
RGB in color  (i.e. in temperature) strongly depends on the
cluster metal content. Indeed, the effective temperature of
a red giant star decreases when the total mass fraction of
heavy elements ($Z$) increases, mainly because of the
larger opacity.  Actually, electron donors, like Fe and
$\alpha$-elements, provides the free electrons needed to
form H$^-$ ions, which are the most important opacity
source in the cool red giant envelope. Thus, stars with
similar $Z$, but different distribution of heavy elements,
should have similar RGB temperatures, if the total amount
of electron donors is similar. This is the case, for
example, of scaled solar and alpha-enhanced mixtures with
the same $Z$ \citep{scs93} {\footnote {The difference in
effective temperature may be somewhat larger at high
metallicity \citep{kim02}.}}. On the contrary, the He
content ($Y$) and the mass ($M$), only slightly affect the
RGB effective temperature:  a similar variation of the
effective temperature  $\delta LogT_e\approx -0.005$  is
indeed obtained for masses increasing from 0.8 to 0.9
M$_\odot$ or for Y increasing from 0.245 to 0.30.  Hence,
the correct parameterization of the RGB location does
require the precise knowledge of the so\--called global
metallicity, which takes also into account the contribution
of the ${\alpha}$--elements, in particular the $[Mg+Si+Fe]$
abundance mixture,  rather than relying on the [Fe/H]
abundance alone \citep{sc91,salc96}. 

In our previous
works \citep{f00,v04,v04a,v05} we have computed the global metallicity
[M/H] from the \cite{cg97} scale ([Fe/H]$_{CG97}$) by adopting an enhancement 
factor of the ${\alpha}$--element linearly decreasing to zero for metal--rich 
clusters with [Fe/H]$_{CG97}>-1$. However, there is now a growing number 
of evidences that this trend is not applicable to the Bulge clusters. In fact,
the most recent high resolution spectroscopic observations of both Bulge 
cluster and field giants \citep{rmw00,c01,ori02,ori04,zoc04,ori05,mike,ori05b}
suggest an ${\alpha}$--enhancement up to solar 
metallicity. Hence, here we have adopted a constant ${\alpha}$--enhancement 
([${\alpha}$/Fe]${\sim}$0.3~dex) over the entire 
-2.2$<$[Fe/H]$_{CG97}<$-0.1 range of metallicities.
For the NGC~6553 and NGC~6528 clusters, which represent 
the metal--rich extremes of our entire database, we 
use the updated values inferred from 
high--resolution IR spectroscopy by \citet{ori02,ori05}
\citep[see also][]{cg97,c01,mel03,zoc04}. 
The adopted metallicity for all the programme clusters are listed in
column~[3] of Table~\ref{tab}.
 
\section{Results and Discussion}

\begin{figure}   
\plotone{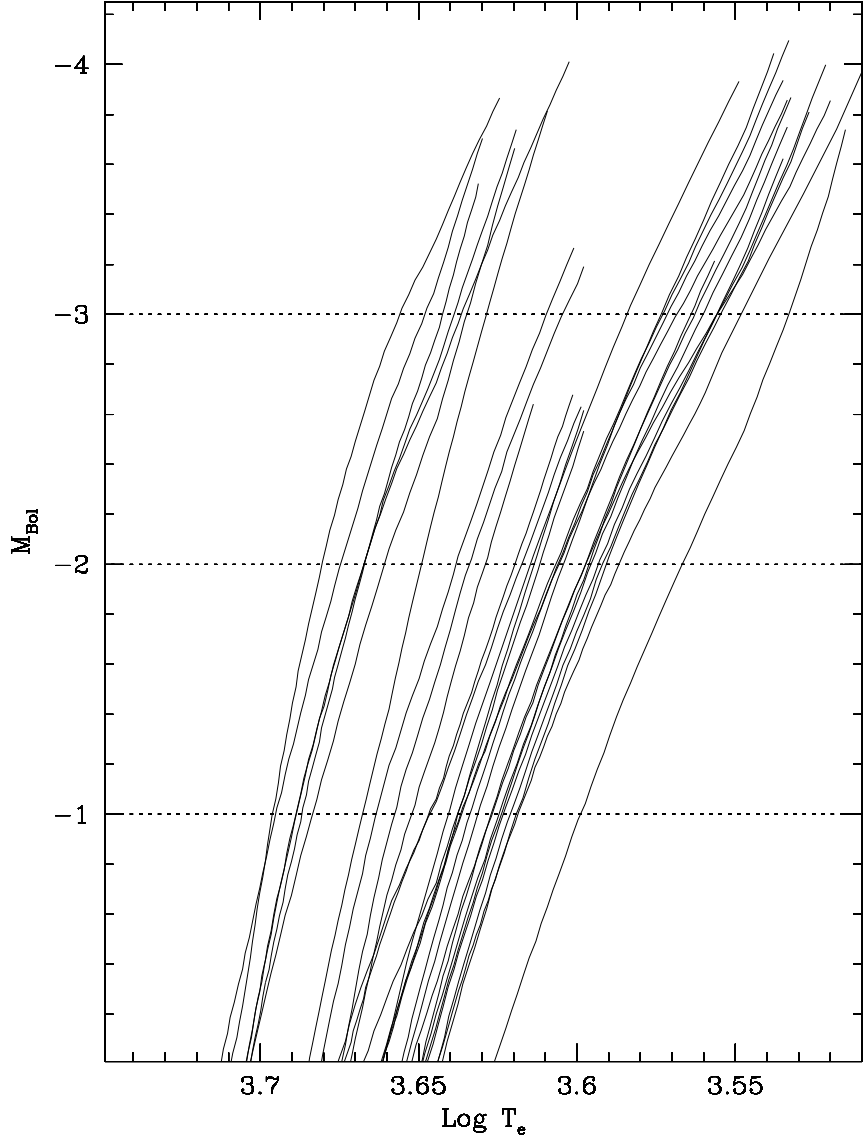}   
\caption{RGB fiducial ridge lines in the [M$_{Bol}$, log(T$_e$)] theoretical
plane for the 28 GGCs in our sample.}
\label{lines} 
\end{figure}

\begin{figure}   
\plotone{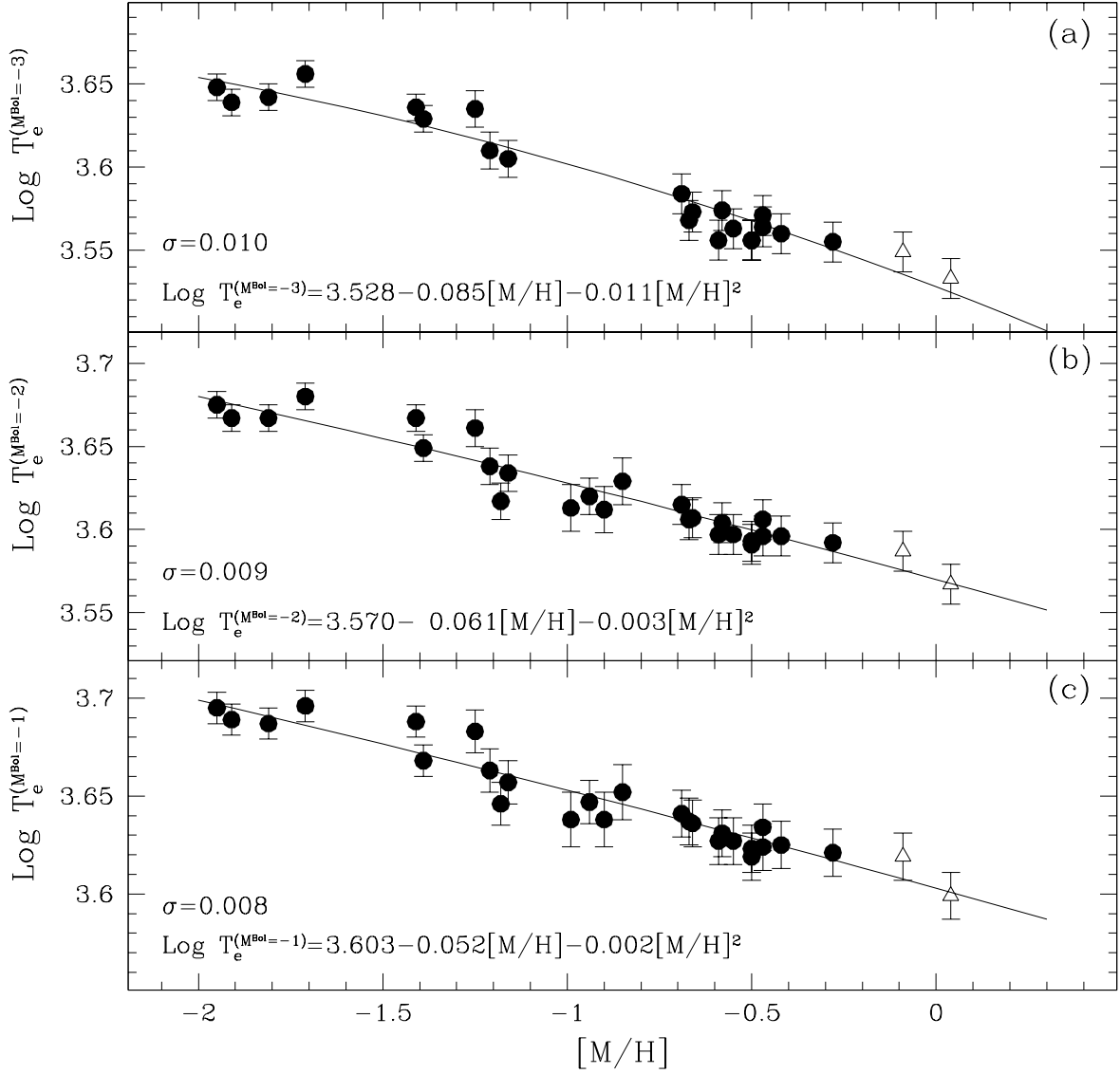}   
\caption{Log T$_e$ at M$_{Bol}$=-3,-2,-1 (panel a, b, c, respectively) as a
function of the global metallicity scale for the 28 GCs in our sample. The solid
lines are the best fit to the data. The empty triangles refer to NGC~6553 and
NGC~6528 with the most recent metallicity estimates by \cite{ori02,ori05} 
respectively.}
\label{temp} 
\end{figure}

We used the CMD plotted in Fig.~\ref{lines} to measure the
RGB effective temperatures at fixed bolometric magnitudes,
namely  M$_{Bol}$=-3, -2, -1, respectively.  The derived
values are listed in Table~\ref{tab} together with a formal
uncertainty of 50--100~K,  and plotted in Fig.~\ref{temp}
as a function of the cluster global metallicity ([M/H]). 
Well defined quadratic relations best fit the observed data
with a very small  r.m.s. ($\rm \Delta log T_e\le0.01$). 
These relations can serve as calibrators for the current
and future generations  of theoretical models for
Population II stars. In the following, we use them  to
calibrate the $\alpha$ parameter for the stellar evolution
code described by \citet[][ hereafter, SCL97]{scl97}, which  adopts the latest
input physics and   the opacity from OPAL above
$logT_{eff}=3.75$, and \citet{fer94} for
$logT_{eff}<3.75$. 

As a first step, we have used this code to
compute a standard solar model and tune the
$\alpha$ parameter in order to reproduce the
solar radius. In doing this we followed the
procedure described in \cite{css95}.  
 
An important consideration concerning the adopted solar
abundances is worth noticing.  
The
\citet[][ hereafter AG89]{ag89} values (giving 
a total metallicity $(Z/X)_\odot=0.0275$) have been widely used for years.
However, more recently significantly lower abundances have 
been proposed by  \citet[][ hereafter, L03]{lod03} 
($(Z/X)_\odot=0.0177$) and by  \citet{asp05}
($(Z/X)_\odot=0.0165$), both using  C,N,O abundances from
3D model atmospheres \citep[see also][]{all02}.  
By adopting both the AG89 and the L03 solar
compositions, we have found that the best SCL97 models to
reproduce   the solar radius require  $\alpha=2.17$  and
$\alpha=1.86$, respectively.

Then, by using the same code (SCL97) and the same
input physics, we have computed a set of RGB models for
low-mass Population II stars. Of course, 
in order to compare the empirical effective
temperatures with those predicted by stellar models, we
need the absolute abundance of metals, namely $Z$. The
relation between these two quantities is:

$$[M/H]=log(Z/X)-log(Z/X)_\odot.$$

Here $X=1-Y-Z$ is the mass fraction of  H in the envelope
of our star/model. Thus, the absolute amount of heavy
elements in a star with known [M/H], substantially depends
on the $Z/X$ ratio in the solar photosphere
[$(Z/X)_{\odot}$].  Hence, for example, for
$[M/H]=-1.5$, one obtains $Z/X$=0.00056, with the latest
solar abundances by L03, or $Z/X$=0.00087, with the widely
adopted AG89 ones. 

   
We have computed a set of RGB models with fixed mass (M=0.8
$M_\odot$). Z has been varied from 0.0001 to 0.02. 
In our analysis, if not explicitly specified,
we adopt $Y=0.245$, a value in agreement with the latest
determination of the primordial He abundance \citep{csi03}.
A $10-20$\% uncertainty in $X$ (or $Y$) induces a variation
of $Z$ that is always smaller than the 
$1\sigma$ error in the $[M/H]$ empirical estimates.
The
initial heavy elements distribution is scaled solar, using 
both the AG89 and L03 references.

\begin{figure}   
\plotone{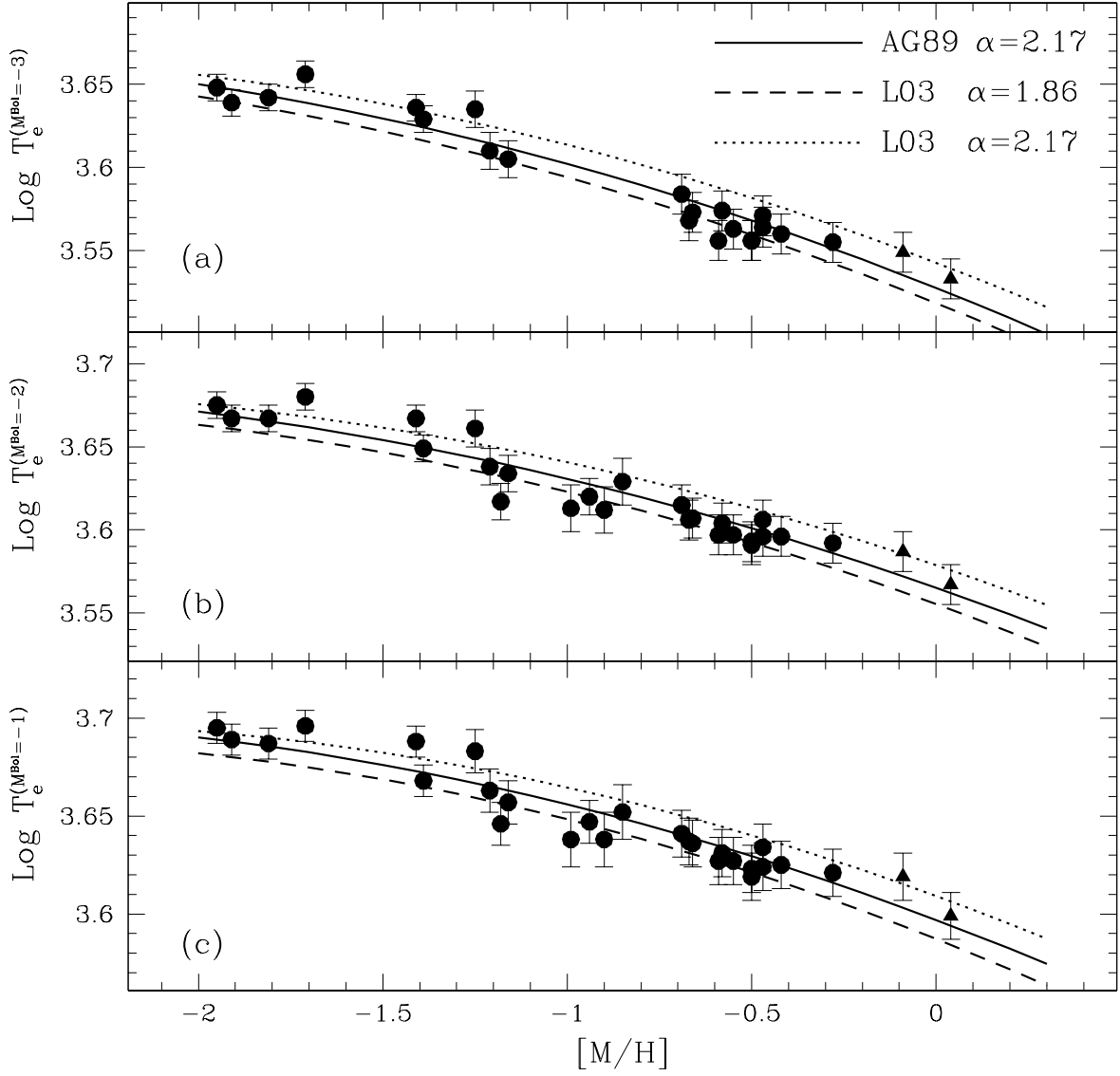}   
\caption{As Fig.\ref{temp} but with overimposed the theoretical predictions:
Model prediction with AG89 solar reference and $\alpha$=2.17
{\it (solid lines)}. Model prediction with L03 solar reference
and $\alpha$=1.86 {\it (dashed lines)} and $\alpha$=2.17
{\it (dotted lines)}.}
\label{theo} 
\end{figure}

Fig.~\ref{theo} reports the results. By using the AG89
solar reference, the measured RGB temperatures are 
well reproduced by the  $\alpha=2.17$ curve, the same value
obtained by best fitting the solar radius. Conversely, by
using the L03 solar reference, the  $\alpha$ value adopted to
reproduce the Sun ($\alpha=1.86$) 
systematically underestimates  the giant temperatures, 
while the model with $\alpha=2.17$  systematically 
overestimates them. The accurate inspection of 
Fig.3 suggests that the observational data are best reproduced by 
a model with an intermediate value of the ML parameter
($\alpha\approx2$).

Hence, from the comparison of the empirical data with 
theoretical predictions, we find that
by adopting the  "old" solar abundances by AG89
both the solar and the Population II Red Giant structure
(regardless of the metallicity) can be nicely reproduced
with a single value of the ML parameter ($\alpha=2.17$).
Conversely, different values of  $\alpha$ for the 
Sun (1.86) and the giant stars (2.0) are required when the L03 solar 
abundances are used. 
Note that the curve with $\alpha=2.0$ for the L03
solar metallicity is not plotted in Figure 3 since it is
nearly coincident with the one obtained with $\alpha=1.86$ 
for the  AG89
solar reference (solid line in Figure 3).

As already stated in Sect.~\ref{intro}, in principle, there is no
theoretical reason that the same value of $\alpha$
should apply to any star, hence,  the two results
are formally equally possible. 

However, there are other issues  
that should be considered and discussed. 
For example,  it must be noticed that 
the empirical abundances for the clusters listed in Table 1,
have been derived using 1D model atmospheres 
and 1D solar references like AG89 or subsequent updates. 
Conversely, the L03 solar references are partially based on 
abundances derived from 3D models. 
Hence, the results using the L03 solar references 
have to be taken with caution, 
since only when   
both a complete set of solar abundances  
and a re-analysis 
of the empirical stellar abundances, fully based on 3D model atmospheres 
\citep[see e.g.][]{kuc05} will be available, firm 
conclusions can be drawn. 

A significant discrepancy between theory and observation
has been already noticed when the "new" L03 
solar abundance have been used to compute standard solar models. 
Indeed, 
the predicted depth of the solar convective zone, as
obtained by adopting the new CNO abundances from \citet{asp03},
is significantly larger than the one derived from
the analysis of the helio-seismic data \citep{bahc03}.  
The bad news is even worst, because an excellent
reproduction of the seismic data was previously found by
adopting the "old" AG89 solar abundances. \citet{bahc05}
invokes a significant increase of the solar Ne (about $10^3$ times larger)
to solve the solar convective zone problem. A recent analysis of X-ray spectra 
of nearby stars \citep{drake05} seems to confirm this expectation. 
Note that such a substantial increase of the solar Ne abundance would also affect the
relation between [M/H] and Z and, in turn, the calibration of the ML.   

While the issue of the firm determination of the
absolute solar abundance will be addressed by future
spectroscopic works, the major result presented here is that,
regardless of the adopted solar reference,   there is not
any clear evidence, within the errors, of a
significant dependence of the ML parameter from
the stellar metallicity \citep[see also][]{p8}.  This is a somewhat
surprising result, since the classical formulation of the
ML theory (Bohm-Vitense 1958)  is a quite naive
approximation of the mean free path of the convettive
bubbles based on very simple assumptions.
The results showed here suggest that in spite of these
crude approximations and the very basic assumptions, the
efficiency of the convective energy transport parametrized
by  $\alpha$ is mainly controlled by the local value of
$H_p$ and it does not require any extra dependence  from
the metallicity of the environment, once an appropriate set
of opacity coefficients have taken into account in the
calculation of the stellar model.

\acknowledgements  
The financial support by the Ministero dell'Istruzione, Universit\'a e Ricerca 
(MIUR) is kindly acknowledged.
This publication makes use of data products from the Two Micron All Sky Survey,
which is a joint project of the University of Massachusetts and Infrared
Processing and Analysis Center/California Institute of Technology, founded by
the National Aeronautics and Space Administration and the National Science
Foundation.

\begin{table}
\scriptsize
\begin{center}
\caption{The cluster sample}
\label{tab}
\begin{tabular}{lcccccc}
\hline
\hline
 & & & &\\
Name  & [Fe/H]$_{CG97}$ & [M/H] & log~T$_e^{(M_{Bol}=-3)}$ & 
log~T$_e^{(M_{Bol}=-2)}$ & log~T$_e^{(M_{Bol}=-1)}$&Ref$^{(*)}$ \\
 & & & &&\\
\hline
 & & & &&\\
M~92  &  -2.16 &  -1.95 &  3.648${\pm}$0.008 & 3.675${\pm}$0.008 &
3.695${\pm}$0.008& V04\\ 
M~15  &  -2.12 &  -1.91 &  3.639${\pm}$0.008 & 3.667${\pm}$0.008 &
 3.689${\pm}$0.008&V04a\\         
M~68  &  -1.99 &  -1.81 &  3.642${\pm}$0.008 & 3.667${\pm}$0.008 &
 3.687${\pm}$0.008& F00\\         
M~30  &  -1.91 &  -1.71 &  3.656${\pm}$0.008 & 3.680${\pm}$0.008 &
 3.696${\pm}$0.008&V04a\\         
M~55  &  -1.61 &  -1.41 &  3.636${\pm}$0.008 & 3.667${\pm}$0.008 &
 3.688${\pm}$0.008& F00\\         
${\omega}$ Cen & -1.60 &  -1.39 &  3.629${\pm}$ 0.008 & 
3.649${\pm}$ 0.008 & 3.668${\pm}$ 0.008&S04$^{(**)}$ \\     
NGC6752& -1.42 &  -1.21&   3.610${\pm}$ 0.011&  3.638${\pm}$ 0.011&
  3.663${\pm}$0.011&V04a\\     
M~10  &  -1.41 &  -1.25&   3.635${\pm}$ 0.011&  3.661${\pm}$ 0.011&
  3.683${\pm}$0.011& V04\\     
M~13 &   -1.39 &  -1.18&   \-------&  3.617${\pm}$ 0.011&
  3.646${\pm}$0.011&V04\\     
M~3 &    -1.34 &  -1.16&   3.605${\pm}$ 0.011&  3.634${\pm}$ 0.011&
  3.657${\pm}$0.011&V04\\     
M~4 &    -1.19 &  -0.94&   \-------&  3.620${\pm}$ 0.011&
  3.647${\pm}$0.011& F00\\     
NGC~362  &  -1.15  & -0.99 &  \------- & 3.613${\pm}$ 0.014 &
 3.638${\pm}$ 0.014&V04a \\   
M~5  &   -1.11 &  -0.90  & \------- & 3.612${\pm}$ 0.014 &
 3.638${\pm}$ 0.014&V04\\     
NGC~288 &   -1.07 &  -0.85 &  \------- & 3.629${\pm}$ 0.014 &
 3.652${\pm}$ 0.014&V04a\\     
NGC~6638 &  -0.97 &  -0.69 &  3.584${\pm}$ 0.012 & 3.615${\pm}$ 0.012 &
 3.641${\pm}$ 0.012&V05\\      
M~107 &  -0.87 &  -0.67 &  3.568${\pm}$ 0.012 & 3.606${\pm}$ 0.012 &
 3.637${\pm}$ 0.012& F00\\      
NGC~6380 &  -0.87  & -0.66 &  3.573${\pm}$0.012 & 3.607${\pm}$0.012 &
 3.636${\pm}$0.012&V04a \\ 
NGC~6569 &  -0.79  & -0.58 &  3.574${\pm}$0.012 & 3.604${\pm}$0.012 &
 3.631${\pm}$0.012&V05 \\ 
NGC~6539 &  -0.71  & -0.50 &  3.556${\pm}$0.012 & 3.593${\pm}$0.012 &
 3.623${\pm}$0.012&O05 \\ 
NGC~6342 &  -0.71  & -0.50 &  3.556${\pm}$0.012 & 3.591${\pm}$0.012 &
 3.619${\pm}$0.012&V04a \\ 
47~tuc & -0.70 &   -0.59  &  3.556${\pm}$0.012 &  3.597${\pm}$0.012 &
 3.627${\pm}$0.012& F00 \\  
NGC~6637 &  -0.68 &  -0.55  & 3.563${\pm}$0.012 & 3.597${\pm}$0.012 &
 3.627${\pm}$0.012&V05 \\ 
NGC~6304 &  -0.68 &  -0.47  & 3.564${\pm}$0.012 & 3.596${\pm}$0.012 &
 3.624${\pm}$0.012&V05 \\ 
NGC~6441 &  -0.68 &  -0.47  & 3.571${\pm}$0.012 & 3.606${\pm}$0.012 &
 3.634${\pm}$0.012&V04a \\ 
NGC~6624 &  -0.63 &  -0.42  & 3.560${\pm}$0.012 & 3.596${\pm}$0.012 &
 3.625${\pm}$0.012&V04a \\ 
NGC~6440 &  -0.49 &  -0.28  & 3.555${\pm}$0.012 & 3.592${\pm}$0.012 &
 3.621${\pm}$0.012&V04a \\ 
NGC~6553 &  -0.30$^{(***)}$ &  -0.09  & 3.549${\pm}$0.012 & 3.587${\pm}$0.012 &
 3.619${\pm}$0.012& F00 \\ 
NGC~6528 &  -0.17$^{(***)}$ &  +0.04  & 3.533${\pm}$0.012 & 3.567${\pm}$0.012 &
 3.599${\pm}$0.012& F00 \\ 
& & && &&\\       
\hline
\end{tabular}
\end{center}
Notes:\\
$^{(*)}$ F00:~\cite{f00}; V04:~\cite{v04}; V04a:~\cite{v04a};
S04:~\cite{bb04}; V05:~\cite{v05}; O05:~\cite{ori05b}\\
$^{(**)}$ The listed values refer to the metal poor
dominant population of $\omega$ Centauri (see S04 for
additional details).\\
$^{(***)}$ From \citet{ori02,ori05}. 
\end{table}

\end{document}